\documentclass[usenatbib]{mn2e}
\usepackage{epsfig}
\usepackage{longtable}
\usepackage{graphicx}
\usepackage{epstopdf}

\def\eps@scaling{.95}
\def\epsscale#1{\gdef\eps@scaling{#1}}

\def\plotone#1{\centering \leavevmode
    \epsfxsize=\eps@scaling\columnwidth \epsfbox{#1}}

\begin{document}

\title[Search for GRB Variability--Afterglow Onset Correlation]{A Search for Correlations between Gamma-Ray Burst Variability and Afterglow Onset}
\author[Yost \& Moore]{S. A. Yost$^1$ and T. M. Moore\\
$^1$College of St. Benedict / St John's University, Collegeville, MN, 56321\\
$^2$SGI, Eagan, MN, 55121}

\maketitle

\begin{abstract}

We compared the time (or time limit) of onset for optical afterglow emission to the $\gamma$-ray variability $V$ in 76 GRBs with redshifts. In the subset (25 cases) with the rise evident in the data, we fit the shape of the onset peak as well and compared the rising and decaying indices to $V$.  We did not find any evidence for any patterns between these properties and there is no statistical support for any correlations. This indicates a lack of connection between irregularities of the prompt $\gamma$-ray emission and the establishment of the afterglow phase. In the ordinary prompt internal shocks interpretation, this would indicate a lack of relationship between $V$ and the bulk Lorentz factor of the event.

\end{abstract}


\section{Introduction}

Gamma-ray bursts (GRBs) are the most luminous explosions in the universe. The $\gamma$-ray dominated prompt emission typically lasts from a fraction of a second to tens of seconds, followed by lower-energy, longer-lasting afterglow. These events are interesting because they are extreme; they require highly relativistic outflows and are associated with the deaths of massive stars or the merger of compact remnants. GRBs can also probe the environment in distant galaxies since the material swept up by the event's external shock drives the afterglow. 

\citet{2006RPPh...69.2259M} gives an exhaustive review of observations and theory as of a few years ago. The broad features of afterglow emission are well described by synchrotron radiation from relativistic electrons swept up by a self-similar external shock, even though the details of the shock such as its microphysics are not always clear. GRB prompt emission is highly irregular compared to the afterglow. Even the prompt radiation mechanism is poorly understood, with models that include internal shocks in the relativistic outflow with synchrotron, inverse Compton radiation or ``jitter'' radiation, as well as magnetic reconnection to dissipate energy, or even radiation from the relativistic hadrons. \citet{2013FrPhy.tmp....2G} and \citet{2012RAA....12.1139M} are recent reviews with more details about theoretical models, including these prompt emission models, in light of ongoing progress from {\em Swift} and {\em Fermi} results.

The prompt mechanisms produce great case-by-case variations in $\gamma$-ray light-curves, from single pulses to very irregular multi-peaked cases or events with pauses between $\gamma$-ray activity. There are also different possibilities for the start of the afterglow, with some initiated during the prompt emission, others delayed until after the high-energy light has ceased, and some showing flares or rebrightening. These effects are not just observed in the X-rays which would have some of the end of the prompt emission as well as afterglow light, but are all seen in optical light (discussed in reviews and notable in compilations of afterglow light-curves such as \citet{2009ApJ...690..163R, 2009ApJ...702..489R}).

Despite many efforts to categorize or correlate prompt $\gamma$-ray properties with other information \citep[e.g., reviewed in the parameter correlation section of][]{2015PhR...561....1K}, there have been few results compared to the understanding offered by afterglows. One area in which a prompt properties could shed light on the afterglow, however, is this range in afterglow onset characteristics. Even the simplest theoretical examination of GRBs expected more than one type of afterglow onset, depending upon properties that would also relate to prompt emission. \citet{1999ApJ...520..641S} noted two regimes of thickness for a relativistic shell whose internal shocks produce the prompt GRB. The division between ``thick'' and ``thin'' depended not only on the shell's physical thickness but also its Lorentz factor. Thin shells were expected to finish emitting prompt radiation before sweeping up enough material to become self-similar and show afterglow while thick shells would have the same timescale to complete the internal shocks (end the GRB) and initiate the afterglow. This would be observed as a rise, or as the end of of dominance by rapidly-decaying reverse shock radiation.

This simple picture does not explain the observed early afterglows, which includes events with established afterglows during the prompt GRB. As seen in afterglow compilations, few events show the characteristic rapidly-declining reverse shock before the slowly-declining synchrotron afterglow. Many events have decaying afterglows during the $\gamma$-ray emission which cannot be attributed to the reverse shock; the rise may be seen or the decay is shallow and continues to late times. Onsets after the GRB have varied optical delays, and may indicate that the events (usually ``long'' bursts) are ``thin'' shell cases \citep[e.g. as applied by][]{2007A&A...469L..13M} and the delay time indicates the initial Lorentz factor of the event in a manner that is only weakly dependent on other factors like the kinetic energy or circumburst density \citep[noted in][]{1999ApJ...520..641S}. However, there are also other models for the delays, such as the effects of off-axis viewing of the relativistic outflow \citep{2013MNRAS.433..759P}.
Moreover, the models are incomplete in other ways. Some events  have observed onsets which are so steep that it is difficult for any model to account for the rapid rise in optical flux \citep[e.g. GRB110205A, where ][ examined  the afterglow for forward and reverse shock onset behaviour and could only match the event to one model by changing the reference time for the start of emission]{2011ApJ...743..154C}.

The range in delays before the self-similar optical afterglow phase continues to be poorly understood. The properties of the afterglow onset have been investigated to compare it to afterglow models, both within the optical itself and between the early optical and X-ray light-curves \citep[with a great deal of work by][]{2010ApJ...725.2209L, 2013ApJ...774...13L}. \citet{2013ApJ...774...13L} has also systematically compared the optical onset lightcurve's peak properties to those of any late optical re-brightening ``bump'' and the event's $\gamma$-ray energy properties. They found correlations between optical properties of the onset peak, and between the optical peak luminosity and the total $\gamma$-ray energy.


This work probes whether there is any connection between measures of optical afterglow onset properties and the ``variability measure'' ($V$) of the prompt GRB. Variability is defined as the normalized squared difference between a GRB light-curve and an appropriately-smoothed version of the light-curve. It should quantify how irregular, spiky, multi-peaked, etc. a burst is. The diversity in this property is obvious among GRB light-curves \citep[noted from early on, e.g.,][]{1995ARA&A..33..415F}, and must connect to some variable physical property that results in irregularity in the prompt emission process(es). 

The variability may also be related to the luminosity of the prompt $\gamma$-rays, e.g. as studied by  \citet{2001ApJ...552...57R}, \citet{2007ApJ...660...16S}, \citet{2009ApJ...707..387X} and others. The irregularity in prompt energy production and emission processes (or the luminosity) could connect to how the event establishes the external shock. Therefore it is particularly interesting whether $V$ connects to the initial optical peak time which should indicate the start of the afterglow phase.

Another possible connection involves afterglow onset delays after the end of a GRB. If this  does indicate that the event has a ``thin'' shell, the timing is mostly dependent upon the initial Lorentz factor. Then a connection between variability and onset time  would indicate a connection between the overall speed of the relativistic outflow and its irregularity. If the observed onset delays are due to another cause like off-axis viewing angles, any connection to $V$ would indicate that the outflow properties related to irregular prompt processes could vary with the off-axis angle of the flow.  

A connection between $\gamma$-ray properties and delays in cases where the afterglow onset is before the end of the GRB is also interesting. This is due to afterglow onsets during a GRB not being expected in the simple GRB picture. \citet{1999ApJ...520..641S} find that in the internal-external shock model, most of the energy of a ``thick'' shell is extracted to start the afterglow at approximately the time it takes to complete the internal shock phase. This is found with simple scalings rather than examining details of how prompt emission might affect the developing external shock. In the case where the afterglow is established during the GRB itself, if the timing is connected to $V$ then it would indicate that highly irregular prompt processes enhance or inhibit the start of the afterglow. This would be a clue to where the simple picture breaks down.

\section{Methods}

We selected a set of GRBs with early observations, complete through early 2012. The definition of ``early'' observations required either a report of a rising optical light-curve, or a detection with some indication of fading behaviour in the first 20 minutes (just over 1 ksec) after the trigger. The set only included those with a redshift measurement, as $z$ has a slight effect on the calculation of $V$ and it is essential to compare items in the GRB emission frame. The set was further limited to {\em Swift}-detected GRBs, as their $\gamma$-ray light-curves and background are consistently available. Finally, a GRB needed a duration measure to be included. We use the $T_{90}$ as a proxy. This is the time to observe 90\% of the fluence above the background starting from when 5\% has been observed, a measure that is nearly always available.\footnote{Rare exceptions when spacecraft pointing prevented a clear view of the end of a GRB produce $T_{90}$ limits, such as GRB080319B, which was not considered.} The final set of 76 events is included with its $V$ calculations in Table \ref{tab_gprops}.

\subsection{Gamma-Ray Variability $V$}\label{s_grbv}

There are many ways to implement the basic description of $V$, especially when attempting to decide what the appropriate smoothing timescale would be and how to minimize dependences on redshift or burst spectral shape (which may not be measurable) or biases from observer effects like time bins. For GRBs, this has led to several different measures \citep[e.g.][]{2000astro.ph..4176F, 2001ApJ...552...57R,2007ApJ...660...16S}. We use the variability equations from \citet{2001ApJ...552...57R}, which takes a smoothing timescale definition that isn't biased by precursors or episodes of low counts. 

\citet{2001ApJ...552...57R} includes two variability measurements, the comparison between the data and the smoothed light-curve (``$V_1$'', the paper's equation 5) and one which approximates the removal of Poisson noise effects on the variability (``$V_2$'', the paper's equation 7). $V_2$ is theoretically a better measure of GRB $\gamma$-ray properties, but its use had problems. $V_1$ is reported as $V$  in table \ref{tab_gprops} and used for variabilities. Values of $V_2$ were also used in the analyses, with no difference in the conclusions.

These problems with $V_2$ may be due to the need to estimate the background as constant, as described below. $V_2$ broadly tracked $V_1$ except for 7 cases where $V_1$ was at the large end and the estimated uncertainty in $V_2$ (equation 8 in the paper) was greater than 1 variability unit (although this was still only about a 10\% uncertainty). Results with $V_1$ did not change when these 7 events were excluded. In 2 of these cases, the denominator for $V_2$ was negative, giving a negative variability value. The denominator is more dependent on the background values than the numerator; this may indicate a significant inaccuracy approximating Poisson noise effects with an approximate, constant background level.

$V$ calculations implement a smoothing timescale that varies between events, using a time required for the detection of a fraction of the fluence. The fraction $f=0.45$ was adopted since that had been used with reasonable results in \citet{2001ApJ...552...57R}. 

The calculations require a time range that includes the complete event, which can be difficult to determine precisely with the noise relative to background. Since the light-curves are of various durations, we calculated $V$ using data that is expected to include the entire GRB time by taking extra time before and after the reported GRB duration. The most common duration measure, $T_{90}$, by design does not include the entire GRB. Rather than use ad hoc times by inspection, we calculated all the measures based on the $T_{90}$ plus a fraction of $T_{90}$ before the trigger and after the end of $T_{90}$. The time range used for the reported values is ``$T_{90}\pm\,30$\%''. However, we also ran ``$\pm\,10$\%'' and ``$\pm\,50$\%'' cases, with no change to our conclusions\footnote{Some values of $V$ increased with the increasing time range, but this did not affect our analysis of correlation evidence.}.

The equations require the total and background counts in each time bin, which were obtained as a background level and the background-subtracted light-curves for the full energy band (15--350 keV). The {\em Swift} background-subtracted light-curves are available as text files from the ground-analysis site\footnote{BAT TXT links at http://gcn.gsfc.nasa.gov/swift\_gnd\_ana.html}. These are in units of counts $s^{-1}$\,cm$^{-2}$ and are converted to counts using the 64-m$s$ bin size and the {\em Swift} 1400 cm$^2$ effective area\footnote{see http://swift.gsfc.nasa.gov/analysis/bat\_digest.html}. The $\gamma$-ray background level was found in the TDRSS records\footnote{BAT TLC links at http://gcn.gsfc.nasa.gov/swift\_gnd\_ana.html} which include non-background-subtracted images of the light-curves in nearly all cases\footnote{Events 050319, 050401, 060927, 070721B, 080906, and 120119A were eliminated as the non-background-subtracted light-curves were not available.} The background was then approximated as constant throughout the GRB event and its value was found by inspection of the event's images.  The units were also counts $s^{-1}$ and converted to counts in a 64-m$s$ bin. The background level was added to the light-curve for the total counts in the calculations.

We verified that changing the assumed background level by 3--5\% had no appreciable effect on the variability calculation. This was a reasonable estimate of our ability to determine a background level from the TDRSS images (a difference of 200--400  in a typical level of 7500 counts $s^{-1}$, which is a noticeable change on the plot axis). Specifically, we noted whether the results changed relative to the only uncertainty calculation available. This is $\delta\,V$ the estimated uncertainty in ``$V_2$''. When we changed the background by $\pm 3$\%, the values of $V$ changed by $<\, \delta\,V$ in 70\% of cases  and by $<\, 2\,\delta\,V$ in 97\% of cases. When the background changed by 5\%, the changes in $V$ were still not extreme by comparison with the estimated uncertainty: 82\% changed by $<\, 2\,\delta\,V$ and 96\% by $<\, 3\,\delta\,V$.

\subsection{Optical Properties}

Most optical peak time limits came from initial decays evident in light-curves from journal articles or GCN notices; Table \ref{tab_optlim} references the data sources. Usually a trend of many points is noted. In cases of sparse early data we required that two successive points in the same filter show an initial decay. If the initial points were separated by a long time baseline (a factor $> 2$) they were not considered a reliable indication of decay since a sharp rise and rollover can occur on that logarithmic timescale; a subsequent pair (e.g., the second and third points) that were more closely-spaced and showed an early decay would set the peak limit time. In a few cases no early light-curve points are published but there is a GCN message indicating an initial decay which then sets the peak time limit for the event.

Cases with an observed rise were fitted to an onset peak shape function, as done by \citet{2010ApJ...725.2209L} (their equation 1).  The full onset through the initial decay is used except in some cases with very long data gaps or where part of the decaying light-curve was excluded due to signs of further flaring / rebrightening. Rather than trying to determine what constitutes a flare to censor a small part of the light-curve, the dataset is cut off, as noted in Table \ref{tab_optpkfits}. Not all of the events fit well statistically to the formula, but the fits did describe the trend of the onset data; see Figures \ref{f_fit1}, \ref{f_fit2}. The results are in Table \ref{tab_optpkfits}, along with information about each data source. This includes events 080804 and 110801a, whose full datasets were not published and were obtained directly from the ROTSE-III project (Akerlof, priv. comm.).

We noted three cases with appropriate $\gamma$-ray data, $z$, and optical peak information which did not easily match the ``decaying gives a peak limit'' or ``rising / falling allows on onset fit'' cases. These are GRBs 070411, 100418A and 100901A. The light-curves are plotted in Figure \ref{f_except}, showing fairly flat initial light-curves. We considered three ways to interpret these events: (1) exclude them (2) take the onset or peak as the point when the optical rise is complete or (3) take the onset or peak as the point where the decay has begun. Table \ref{tab_optpkspecial} discusses the values for options (2) and (3). We performed calculations for all three methods and the exclusion or inclusion method did not affect the results. The results and calculations presented in this paper simply exclude these three events.

\subsection{Checking for Correlations}\label{endofsec2}

While the optical onset time could have a relationship to the $\gamma$-ray properties, it was also interesting to examine any onset properties including the onset peak shape parameters. There have been some cases of unexpected early afterglow rise and decay rates \citep[discussed in section 6 of][]{2006RPPh...69.2259M}. There are no theories directly expecting the $\gamma$-ray variability to cause these effects, but any correlation could provide a clue to the unusual cases.  

An initial examination versus $V$ showed no evident patterns (see Figures \ref{f_tpkz}--\ref{f_ratio}). The rise-to-decay index ratio $r/d$ might have a loose positive correlation with $V$.  Peak times were examined both  de-redshifted ($T_{pk}/(1+z)$) and as a fraction of the GRB duration proxy $T_{90}$ ($T_{pk}/T_{90}$). The points for the peak times suggest a loose correlation with $V$. The peak limits, however, do fill in much of the parameter space. 

The relationships were then investigated quantitatively via correlation tests using a Spearman rank coefficient $\rho$, as implemented by the Iraf STSDAS package. This calculation permits censored points (limits like the peak times) but does not use uncertainty information.

The Spearman results are in Table \ref{tab_spear}. For 20 cases of optical peak fits, the indices $r$ and $d$ were constrained (reported uncertainty $< 1/3$ of the fitted value), allowing comparisons of $V$ to each index and the ratio $r$/$d$ (a proxy for asymmetry). These are fewer than the threshold of 30 data points for a reliable Spearman correlation test; the results are included to show that there are no statistically obvious correlations which could be hinted at by the results with 20 points. The fitted peak flux density is not considered, as the comparison between events would require accounting for poorly-constrained local extinction along with very careful adjustments for different filters and redshifts.

\section{Results \& Discussion}

\begin{figure}
\epsscale{1.0}
\plotone{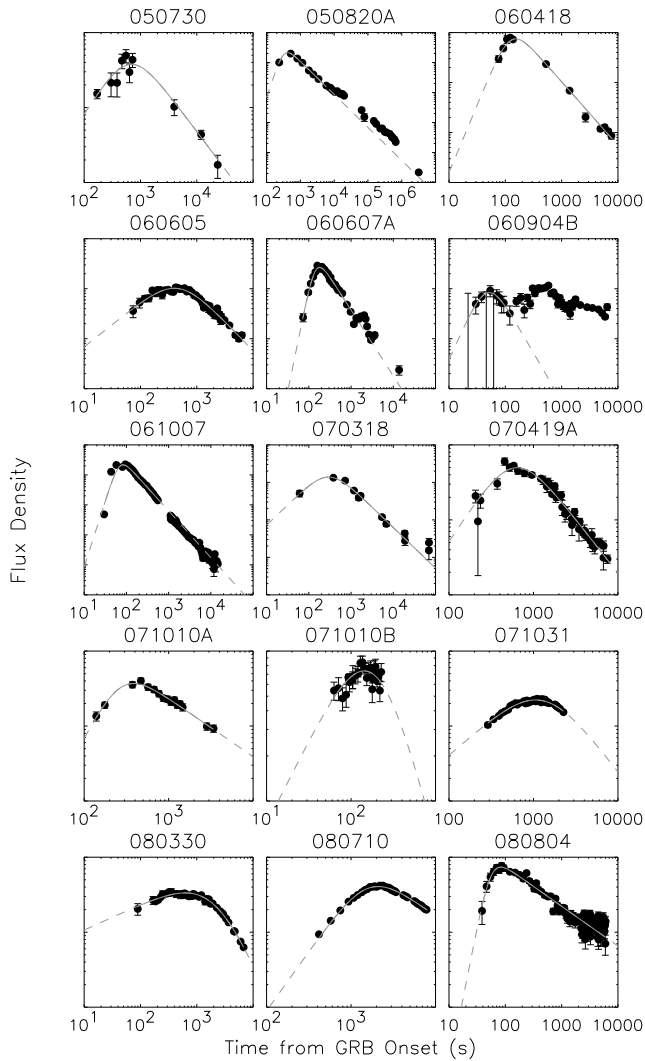}
\caption{Fits to the cases with evidence of a rising peak, part 1. Fit results and data references are in Table \ref{tab_optpkfits}. Later times are excluded in some cases to avoid flares, rebrightenings, steepening decay, or data gaps; these parts are shown in dashed lines for the fit model. The fits give a good representation of the shape even in cases with poor $\chi^2$, and so are useful for determining the peak times for further analysis\label{f_fit1}}
\end{figure}

\begin{figure}
\epsscale{1.0}
\plotone{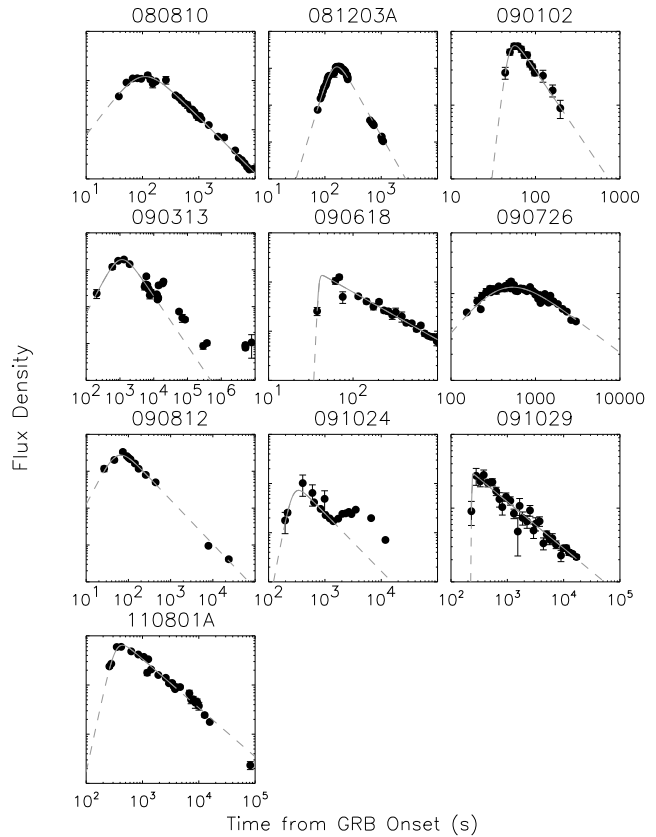}
\caption{Fits to the cases with evidence of a rising peak, part 2. See Figure \ref{f_fit1} for details.\label{f_fit2}}
\end{figure}

\begin{figure}
\epsscale{1.0}
\plotone{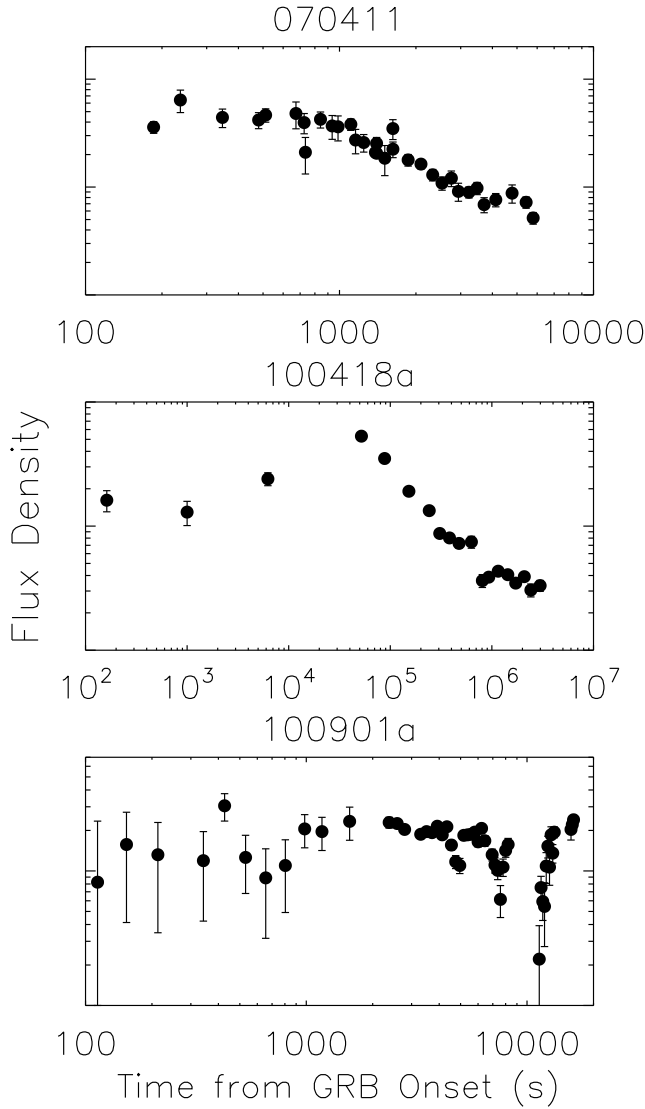}
\caption{``Difficult to Classify'' cases. These neither have a well-defined initial rise and rollover nor an evident initial decay. 070411 is directly adapted from Figure 2 of \citet{2008AIPC.1000..257F}, but does not show the time error bars. These events are treated 3 ways in further analyses: ignored, with the peak set by the end of the rise, and with the peak set by the start of the decay. Further details are in Table \ref{tab_optpkspecial}, which gives the data references.\label{f_except}}
\end{figure}

\begin{figure}
\epsscale{1.0}
\plotone{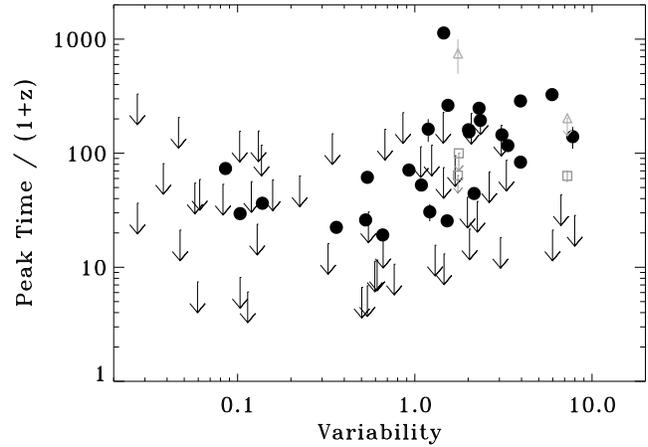}
\caption{Source frame peak time vs $\gamma$-ray variability. Points from Table \ref{tab_optpkfits}, limits from Table \ref{tab_optlim}. The 3 ``hard to classify'' cases are included, with points tagged by grey squares for the peak time set by the end of the rise and grey triangles for the time set by the start of the decay. The  decay-set peak time of 100418A is off the plot. The detected points suggest a loose positive correlation. It is important to include the peak limits, as they fill in significant parts of the parameter space. The Spearman test (Table \ref{tab_spear}) indicates that there is no statistical support for a significant correlation between the properties, with a null probability of 2\%.\label{f_tpkz}}
\end{figure}

\begin{figure}
\epsscale{1.0}
\plotone{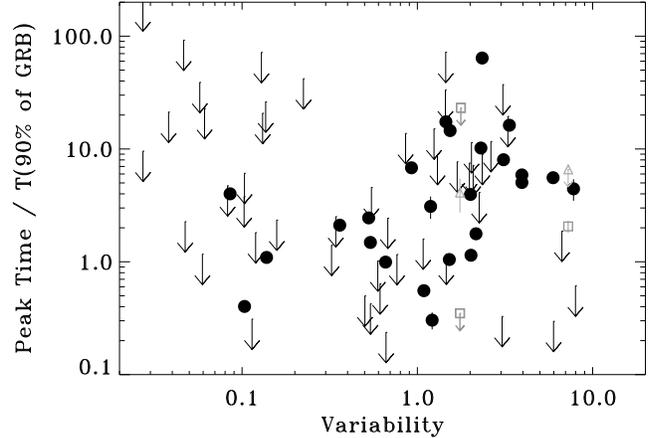}
\caption{Peak time as a fraction of GRB duration vs $\gamma$-ray variability. Points and ``hard to classify'' cases presented as in Figure \ref{f_tpkz}. The fitted peaks (points) alone also suggest a loose positive correlation, and including the peak limits shows that the results fill in more of the parameter space. The Spearman test (Table \ref{tab_spear}) indicates that there is no statistical support for a significant correlation between the properties, with a null probability of 2\% \label{f_tpk90}}
\end{figure}

Figures \ref{f_tpkz} and \ref{f_tpk90} show the variation of peak times (de-redshifted and as a fraction of the GRB duration) versus variability; the points show the uncertainties reported from the fits. With some irregular instead of smooth behaviour, not all the fits are statistically good as shown by the $\chi^2$ in Table \ref{tab_optpkfits}. However, figures \ref{f_fit1} and \ref{f_fit2} demonstrate that the fits match the peak times and overall shapes. While the peak time points suggested a trend, the Spearman coefficients show null (no correlation) probabilities of 1.6\%  for the source frame peak times and 2.0\% for the times as a fraction of the duration $T_{90}$. These confidence levels for a correlation would correspond to 2.4 $\sigma$ and 2.3 $\sigma$ in a normal distribution.

\begin{figure}
\epsscale{1.0}
\plotone{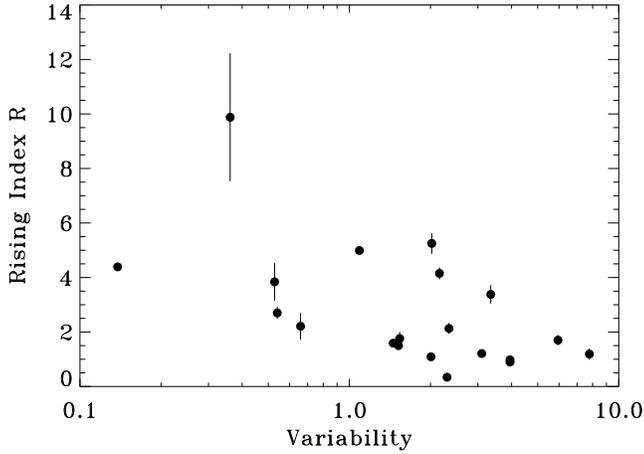}
\caption{Rising index $r$ vs $\gamma$-ray variability $V$.  The indices are from the fit results of  Table \ref{tab_optpkfits}, the asymptotic powerlaw of the fitting shape. Only the 20 events where the formal uncertainty is $<\,1/3$ of $r$ are shown. The plot suggests a loose relationship,  with a lack of high-$r$, high-$V$ events. But there is no statistical support for a connection between the properties, with the Spearman test of Table \ref{tab_spear}  giving a 0.9\% null probability and noting that the 20 events is insufficient to trust the accuracy of the test results. \label{f_rind}}
\end{figure}

Figures \ref{f_rind}--\ref{f_ratio} show the indices and rising/falling index ratio versus variability. The plots of the decaying index and the ratio $r$/$d$ suggest a broad relationship, but again there is no statistical support for it. The null probabilities are 0.9\% for $r$, 45\% for $d$ and 1.6\% for $r/d$, or no better than confidence levels equivalent to  2.6, 0.8, or 2.4 $\sigma$ in a normal distribution.

As previously discussed, there are uncertainties in the treatment of the data (three hard-to-classify cases, other time bases, etc). Some of these provide even larger null probabilities for correlations. Moreover the Spearman test does not take uncertainties into account. These factors tend to lessen any confidence in the slight (and not statistically significant) confidence levels for a correlation.  There is therefore no statistically compelling support for a connection between the variability of the prompt GRB emission and the afterglow onset properties, particularly the optical rise times.


\onecolumn
\begin{center}
\begin{longtable}{lcccc}
\caption{GRBs and their Variability Measures}\label{tab_gprops}\\
\hline
GRB & $ T_{90}\, (s)$ & $ z $ & $ V  $   & $z$ Reference  \\
\hline
050502A & 58.9 & 3.793 & 1.5 & \citet{2005GCN..3332....1P}\\
050525A & 8.8 & 0.606 & 2.0 & \citet{2005GCN..3483....1F}\\
050730 & 156.5 & 3.969 & 7.8 & \citet{2005GCN..3732....1P} \\
050802 & 19 & 1.71 &  1.2 & \citet{2005GCN..3749....1F} \\
050820A & 26 & 2.615 & 3.4 & \citet{2005GCN..3860....1L} \\
050908 & 19.4 & 3.35 &  3.3 & \citet{2005GCN..3948....1F} \\
050922C & 4.5 & 2.198 &  0.057 &\citet{2005GCN..4029....1J} \\
051109A & 37.2 & 2.346 & 0.60 & \citet{2005GCN..4221....1Q} \\
051111 & 46.1 & 1.549 &  0.61 & \citet{2005GCN..4271....1P} \\
060108 & 14.3 & 2.03 &  3.1 & \citet{2006GCN..4539....1M} \\
060206 & 7.6 & 4.045 &  0.22 & \citet{2006GCN..4692....1F} \\
060418 & 103.1 & 1.49 & 0.54 & \citet{2006GCN..5002....1P}  \\
060502A & 28.4 & 1.51 &  0.083 & \citet{2006GCN..5052....1C} \\
060510B & 275.2 & 4.9 &  8.0  & \citet{2006GCN..5104....1P}\\
060512 & 8.5 & 0.4428 &  2.6 &\citet{2006GCN..5217....1B} \\
060605 & 79.1 & 3.8 &  3.9 & \citet{2006GCN..5223....1P} \\
060607A & 102.2 & 3.082 &  2.2 & \citet{2006GCN..5237....1L} \\
060904B & 171.5 & 0.703 &  1.2 & \citet{2006GCN..5513....1F} \\
060908 & 19.3 & 1.884 &  0.55 & \citet{2009ApJS..185..526F} \\
060912 & 5 & 0.937 &  0.061 & \citet{2006GCN..5617....1J} \\
060926 & 8 & 3.208 &  2.0 & \citet{2006GCN..5637....1D} \\
061007 & 75.3 & 1.261 &  0.14 & \citet{2006GCN..5715....1O} \\
061110A & 40.7 & 0.758 &  6.7 & \citet{2007GCN..6759....1F} \\
061110B & 134 & 3.44 &  2.4 & \citet{2006GCN..5809....1F} \\
070318 & 74.6 & 0.836 &  2.0 & \citet{2007GCN..6216....1J} \\
070411 & 121.5 & 2.954 &  7.3 & \citet{2007GCN..6283....1J} \\
070419A & 115.6 & 0.97 &  5.9 & \citet{2007GCN..6322....1C} \\
071003 & 150 & 1.1 &  6.0 & \citet{2007GCN..6850....1P} \\
071010A & 6 & 0.98 &  2.3 & \citet{2007GCN..6864....1P} \\
071010B & 35.7 & 0.947 &  0.086 & \citet{2007GCN..6888....1C} \\
071020 & 4.2 & 2.142 &  0.10 & \citet{2007GCN..6984....1F} \\
071031 & 180 & 2.692 &  3.9 & \citet{2007GCN..7023....1L} \\
080210 & 45 & 2.641 &  1.7 & \citet{2008GCN..7286....1J} \\
080319C & 34 & 1.95 &  0.32 & \citet{2008GCN..7517....1W} \\
080330 & 61 & 1.51 &  2.3 & \citet{2008GCN..7544....1M} \\
080413A & 46 & 2.433 &  0.50 & \citet{2008GCN..7602....1T} \\
080413B & 8 & 1.10 &  0.027 & \citet{2008GCN..7601....1V} \\
080430 & 16.2 & 0.767 &  0.77 & \citet{2008GCN..7654....1C} \\
080603B & 60 & 2.69 &  0.54 & \citet{2008GCN..7797....1F} \\
080605 & 20 & 1.64 &  0.13 & \citet{2008GCN..7832....1J} \\
080607 & 79 & 3.036 &  0.11 & \citet{2008GCN..7849....1P} \\
080710 & 120 & 0.845 &  1.5 & \citet{2008GCN..7962....1P} \\
080721 & 16.2 & 2.602 &  0.14 & \citet{2008GCN..7997....1D} \\
080804 & 34 & 2.2045 &  0.53 & \citet{2008GCN..8058....1T} \\
080805 & 78 & 1.505 &  2.1 & \citet{2008GCN..8077....1J} \\
080810 & 106 & 3.35 &  1.5 & \citet{2008GCN..8083....1P} \\
080913A & 8 & 6.7 &  1.5 & \citet{2008GCN..8225....1F} \\
081029 & 270 & 3.848 &  3.0 & \citet{2008GCN..8438....1D} \\
081203A & 294 & 2.1 & 1.1 & \citet{2008GCN..8601....1L} \\
081222 & 24 & 2.77 &  0.059 & \citet{2008GCN..8713....1C} \\
090102 & 27 & 1.547 &  0.36 & \citet{2009GCN..8766....1D} \\
090313 & 79 & 3.375 &  1.5 & \citet{2009GCN..8994....1C} \\
090424 & 48 & 0.544 &  0.12 & \citet{2009GCN..9243....1C} \\
090426 & 1.2 & 2.609 &  0.13 & \citet{2009GCN..9264....1L} \\
090618 & 113.2 & 0.54 &  0.10 & \citet{2009GCN..9518....1C} \\
090715B & 266 & 3.00 &  0.68 & \citet{2009GCN..9673....1W} \\
090726 & 67 & 2.71 &  3.1 & \citet{2009GCN..9712....1F} \\
090812 & 66.7 & 2.452 &  0.66 & \citet{2009GCN..9771....1D} \\
091018 & 4.4 & 0.971 &  0.046 & \citet{2009GCN..10038...1C} \\
091024 & 109.8 & 1.092 &  1.2 & \citet{2009GCN..10065...1C} \\
091029 & 39.2 & 2.752 &  0.93 & \citet{2009GCN..10100...1C} \\
100316B & 3.8 & 1.180 &  1.3 & \citet{2010GCN..10495...1V} \\
100418A & 7 & 0.6235 &  1.8 & \citet{2010GCN..10620...1A} \\
100621A & 63.6 & 0.542 &  0.10 & \citet{2010GCN..10876...1M} \\
100728A & 198.5 & 1.567 &  0.66 & \citet{2013GCN14500} \\
100814A & 174.5 & 1.44 &  1.1 & \citet{2010GCN..11089...1O} \\
100816A & 2.9 & 0.8034 &  0.027 & \citet{2010GCN..11123...1T} \\
100901A & 439 & 1.408 &  1.8 & \citet{2010GCN..11164...1C} \\
110128A & 30.7 & 2.339 &  2.3 & \citet{2011GCN..11607...1S} \\
110422A & 25.9 & 1.77 &  0.047 & \citet{2011GCN..11977...1M} \\
110503A & 10 & 1.613 &  0.038 & \citet{2011GCN..11993...1D} \\
110801A & 385 & 1.858 &  2.0 & \citet{2011GCN..12234...1C} \\
111107A & 26.6 & 2.893 &  1.5 & \citet{2011GCN..12537...1C} \\
111228A & 101.2 & 0.7163 &  0.34 & \citet{2011GCN..12770...1S} \\
120326A & 69.6 & 1.798 &  0.16 & \citet{2012GCN..13118...1T} \\
120327A & 62.9 & 2.813 &  0.86 & \citet{2012GCN..13146...1S} \\
\hline
\end{longtable}
\medskip
GRB events used are presented with their observed 90\% $\gamma$-ray duration $T_{90}$ (from the {\em Swift} archive) and their redshift. Variability $V$ examines the irregularity of the $\gamma$-rays by comparing the GRB lightcurve relative to itself smoothed by a time corresponding to observing 45\% of the fluence. The variability is the first measure discussed in \citet{2001ApJ...552...57R}. As noted, there were problems with the second measure which  approximately accounts for removing Poisson noise variability effects from this calculation. The results presented use $T_{90}\pm\, 30$\% to get a background region. Variabilities were also calculated using $\pm\,10$\% and $\pm\,50$\%. 
Some numbers changed between background selection regions by more than the estimated variability uncertainty, which is actually the $\delta\,V$ estimate for the closely-related second variability measure of \citet{2001ApJ...552...57R}. However, the overall conclusions about the lack of evidence for correlations did not.
\end{center}
\twocolumn


\begin{table*}
\begin{minipage}{126mm}
\caption{Optical Peak Time Upper Limits}
\label{tab_optlim}
\begin{tabular}{@{}ccc}
\hline
GRB & $ T_{PK} $ Limit (s) & Data Source  \\
\hline
050502A & 68.95 & \citet{2006ApJ...636..959Y} \\
050525A & 66 &  \citet{2009ApJ...702..489R} \\
050802 & 286 & \citet{2007MNRAS.380..270O} \\
050908 & 377 & \citet{2005GCN..3960....1D} \\
050922C & 174.9 & \citet{2009ApJ...702..489R} \\
051109A & 37.9 & \citet{2009ApJ...702..489R} \\
051111 & 29.4 & \citet{2007ApJ...657..925Y} \\
060108 & 532 & \citet{2006MNRAS.372..327O} \\
060206 & 319 & \citet{2006ApJ...648.1125M} \\
060502A & 134 & \citet{2006GCN..5068....1P} \\
060510B & 168 & \citet{2006GCN..5103....1M}  \\
060512 & 99 & \citet{2006GCN..5130....1D} \\
060908 & 88 & \citet{2009MNRAS.395..490O} \\
060912 & 114 & \citet{2009MNRAS.395..490O} \\
060926 & 91 & \citet{2006GCN..5632....1L, 2006GCN..5901....1L} \\
061110A & 76 & \citet{2006GCN..5798....1Z} \\
061110B & 1200 & \citet{2006GCN..5804....1M} \\
071003 & 44.5 & \citet{2008ApJ...688..470P}\\
071020 & 25.6 & \citet{2007GCN..6948....1S, 2007GCN..6951....1Y} \\
080210 & 345.9 & \citet{2008GCN..7280....1K} \\
080319C & 47.6 & \citet{2008GCN..7477....1W} \\
080413A & 22.9 &  \citet{2008GCN..7593....1R} \\
080413B & 76.5 & \citet{2011AnA...526A.113F} \\
080430 & 18.8 & \citet{2008GCN..7646....1K} \\
080603B & 25.4 & \citet{2008GCN..7792....1R} \\
080605 & 414 & \citet{2008GCN..7845....1K} \\
080607 & 24.5 & \citet{2011AJ....141...36P} \\
080721 & 423 & \citet{2009MNRAS.400...90S} \\
080805 & 558 & \citet{2008GCN..8061....1D} \\
080913A & 576 & \citet{2009ApJ...693.1912G} \\
081029 & 88.5 & \citet{2011AIPC.1358..130H} \\
081222 & 28 &  \citet{2008GCN..8692....1C} \\
090424 & 87 & \citet{2009GCN..9225....1X} \\
090426 & 86 & \citet{2011MNRAS.410...27X} \\
090715B & 648 & \citet{2009GCN..9677....1G} \\
091018 & 405.6 & \citet{2012MNRAS.426....2W} \\ 
100316B & 34 & \citet{2010GCN..10494...1H} \\
100621A & 240 & \citet{2010GCN..10874...1U} \\
100728A & 47 & \citet{2010GCN..11007...1P} \\
100814A & 278 & \citet{2010GCNR..303....1S} \\
100816A & 595.7 & \citet{2010GCNR..300....1O} \\
110128A & 126.2 & \citet{2011GCN..11613...1L} \\
110422A & 58.7 & \citet{2013ARep...57..233G} \\
110503A & 212 & \citet{2011GCN..12000...1O} \\
111107A & 885 & \citet{2011GCN..12535...1L, 2011GCN..12544...1L} \\
111228A & 254.1 & \citet{2011GCN..12789...1K} \\
120326A & 163 & \citet{2012GCN..13107...1K, 2012GCN..13108...1K} \\
120327A & 864 & \citet{2012GCN..13144...1M} \\
\hline
\end{tabular}

\medskip
Subsample with early optical or near-IR decays giving limits on the optical peak time, and so the onset of the optical afterglow. Initial decays were noted within 20 minutes of the trigger, thus some optical emission has already risen regardless of subsequent flares or rebrightening. Many references are  from GCN Notices. Usually a trend of several points indicates the decay. Sparse early data required two consecutive points showing decay in the same filter, separated in time by no more than a factor of 2 to avoid the possibility of missing a sharp peak. Cases where the first two points were too far apart in time used later points to set the limit, e.g., the second \& third. In a few cases, a single GCN Notice indicates one point but declares that the transient is fading, which is taken as sufficient confirmation for the limit. As can be seen in Figures \ref{f_tpkz} and \ref{f_tpk90}, these limits fill in part of peak time--variability plot, while the detections alone make these appear to be loosely correlated.
\end{minipage}
\end{table*}


\begin{table*}
\begin{minipage}{126mm}
\caption{Optical Initial Peak Fits}
\label{tab_optpkfits}
\begin{tabular}{@{}cccccccccc}
\hline
GRB & $ T_{PK} $ (s) & $\delta\, T_{PK} $ (s) & $r$ & $\delta\, r$ & $d$ & $\delta\, d$ & $\chi^2$ / DOF & Avoiding flares/dips & Data source \\
\hline
050730 & 690 & 150 & 1.19 & 0.21 & 1.00 & 0.07 & 6.7 / 6 &  & 1  \\
050820A & 422 & 12 & 3.38 & 0.34 & 1.10 & 0.02 & 17.6 / 4 & $t \le 7$ks & 2 \\
060418 & 153.3 & 3.3 & 2.70 & 0.22 & 1.27 & 0.02 & 15.8 / 8 &  & 3  \\
060605 & 399 & 12 & 0.90 & 0.09 & 1.17 & 0.05 & 74.5 / 50 &  & 4 \\
060607A & 180.9 & 2.4 & 4.15 & 0.22 & 1.32 & 0.04 & 45.7 / 23 & $t \le 1$ks & 3 \\
060904B & 52.1 & 8.9 & 2.51 & 2.48 & 1.94 & 2.15 & 2.2 / 7 & $t \le 130$s  & 4 \\
061007 & 83.39 & 0.45 & 4.39 & 0.05 & 1.56 & 0.01 & 2377 / 19 & $t \le 600$s  & 4 \\
070318 & 295 & 21 & 1.09 & 0.14 & 1.08 & 0.05 & 17.2 / 8 &  & 5 \\
070419A & 643 & 20 & 1.70 & 0.19 & 1.44 & 0.05 & 47.5 / 42 & $t \le 10$ks  & 6  \\
071010A & 384 & 22 & 2.13 & 0.20 & 0.78 & 0.05 & 4.8 / 12 & $t \le 3.6$ks  & 7  \\
071010B & 143 & 12 & 1.13 & 0.69 & 2.17 & 5.41 & 17.2 / 24 &  & 8  \\
071031 & 1057 & 29 & 0.98 & 0.05 & 1.48 & 0.19 & 29.4 / 39 & $t \le 2.3$ks  & 9  \\
080330 & 622 & 16 & 0.34 & 0.02 & 1.77 & 0.11 & 30.1 / 36 & $t \le 10$ks  & 10 \\
080710 & 2090.6 & 4.9 & 1.59 & 0.01 & 0.820 & 0.005 & 596 / 36 & $t \le 8$ks  & 11 \\
080804 & 83.3 & 3.9 & 3.84 & 0.69 & 0.53 & 0.01 & 173 / 85 &  & 0 \\
080810 & 111.4 & 1.4 & 1.50 & 0.05 & 1.14 & 0.01 & 132 / 28 &  & 12  \\
081203 & 163.42 & 0.26 & 4.99 & 0.02 & 2.68 & 0.04 & 1037 / 42 & $t \le 400$s  & 13. \\
090102 & 57.0 & 1.5 & 9.88 & 2.35 & 1.69 & 0.18 & 4.9 / 9 &  & 14 \\
090313 & 1154 & 91 & 1.76 & 0.24 & 1.35 & 0.14 & 19.3 / 8 &  & 15 \\
090618 & 45.47 & 0.41 & 24.53 & 8.78 & 0.96 & 0.03 & 26.9 / 18 & $t \le 1$ks & 0 \\
090726 & 538 & 11 & 1.21 & 0.05 & 0.73 & 0.03 & 198 / 43 &  & 16, 17 \\
090812 & 66.3 & 4.0 & 2.21 & 0.49 & 1.20 & 0.12 & 9.7 / 6 & $t \le 1$ks & 18  \\
091024 & 340 & 76.4 & 6.00 & 5.57 & 1.17 & 0.06 & 3.8\footnote{4 early GCN points with no uncertainties were employed with 0.5-magnitude uncertainties} / 6 & $t \le 1.5$ks &  19--24,  \\
091029 & 267.2 & 4.3 & 30.82\footnote{The first two points show a steep rise, but this index $r$ is inconsistent with the index found in the $i$ band, casting doubt on the optical onset shape fit. This event was not used in tests for $r$ and $d$ correlations}& 0.55 & 0.63 & 0.02 & 34.7 / 27 &  & 25 \\
110801A & 439.9 & 8.2 & 5.25 & 0.38 & 0.99 & 0.02 & 105 / 14 & $t \le 4$ks  & 0, 26--28   \\
\hline
\end{tabular}
\medskip
Subsample with an initial rise, fitted to an empirical shape with rising / decaying asymptotic powerlaw indices $r$ and $d$, the peak time $T_{pk}$, and the peak flux density. The peak flux density is not considered for further analysis, as this would require corrections for poorly-known local extinction, so the results are not reported. In some cases later parts of the available data have been excluded from the fits to avoid flares, rebrightening, steepening after the initial decay, or a data gap in the decay; these are noted in the column ``Avoiding flares / dips''. The fits are all plotted in Figures \ref{f_fit1} and \ref{f_fit2}, with dashed lines for any part of the time range which was excluded from the fit. Except for 061007's extremely sharp rise, the fits give a good overall match to the shape of the onset even when the $\chi^2$ is large. The formal uncertainties on the parameters may not account for the full uncertainty when the fit is not good, and the formal uncertainties show that $r$ and $d$ are unconstrained in some cases. In such cases, $T_{pk}$ is useful for correlation analysis but $r$ and $d$ are not.\newline
0. {\em Akerlof, priv. comm.}; 1. \citet{2005GCN..3717....1B}; 
2. \citet{2010ApJ...711..641C};
3. \citet{2007A&A...469L..13M}; 
4. \citet{2009ApJ...702..489R}; 
5. \citet{2009ApJ...690..163R};
6. \citet{ 2009MNRAS.395.1941M};
7. \citet{2008MNRAS.388..347C}; 
8. \citet{2008ApJ...679.1443W}; 
9. \citet{2009ApJ...697..758K}; 
10. \citet{2009A&A...499..439G};
11. \citet{2009A&A...508..593K}; 
12. \citet{2009MNRAS.400..134P}; 
13. \citet{2009MNRAS.395L..21K};  
14.  \citet{2009GCN..8764....1K}; 
15. \citet{2010ApJ...723.1331M};	
16 \citet{2009GCN..9715....1M}; 
17 \citet{2010A&A...510A..49S}; 
18. \citet{2009arXiv0908.2849D}; 
19. \citet{2011A&A...528A..15G};
20. \citet{2009GCN..10063...1M}; 21. \citet{2009GCN..10066...1C}; 22. \citet{2009GCN..10073...1H}; 23. \citet{2009GCN..10074...1U}; 24.\citet{2009GCN..10075...1C}; 25. \citet{2009GCN..10107...1L};26. \citet{2011GCN..12241...1S}; 27. \citet{2011GCN..12238...1P}; 28. \citet{2011GCN..12233...1K};
\end{minipage}
\end{table*}


\begin{table*}
\begin{minipage}{126mm}
\caption{Optical Rise / Peak Special Cases}
\label{tab_optpkspecial} 
\begin{tabular}{@{}cccc}
\hline
GRB & $ T_{PK} $ (s) & $ T_{PK} $ (s) & data source \\
 & established by rise & established by decay & \\
\hline
070411 & $250 \pm 30$ & $<\, 800$ & \citet{2008AIPC.1000..257F} \\
100418A  & $<\, 162 $ & $< 5 \times 10^4$ &  \citet{2011ApJ...727..132M} \\
100901A  & $<\, 153.4$ & $1800 \pm 600$ & \citet{2012MNRAS.421.1874G} \\
\hline
\end{tabular}

\medskip
Three events with marginal evidence of a rise, or an apparent plateau. They are shown in Figure \ref{f_except}; 070411's data is adapted from Figure 2 of \citet{2011ApJ...727..132M}, but does not show the time error bars. The events are treated 3 different ways for the search for correlations between $\gamma$-ray variability and the optical onset. They are given a peak value or limit consistent with ``onset'' being at the end of an optical rise (limited by being before a plateau), before the beginning of a decay (conservatively limited by the end of a plateau), or excluded from the analysis. 070411 is assigned a peak value instead of a limit for the first case, due to the discussion of a rising phase in \citet{2011ApJ...727..132M}; the uncertainty comes from the duration of the second light-curve point. 100901A is assigned a peak value for the second case as the optical peak fit was attemped through the clear decay after 1 hour.  The results were not constrained and are not in Table \ref{tab_optpkfits}, but this gives a good estimate of the start of the decay.
\end{minipage}
\end{table*}


\begin{table*}
\begin{minipage}{126mm}
\caption{Spearman Correlation Coefficient Results}
\label{tab_spear}
\begin{tabular}{@{}ccc}
\hline
Item Compared & Spearman $\rho$ & Null Probability \\
\hline
 $T_{pk}/T_{90}$ (1) &   0.274 & 0.0201 \\
 $T_{pk}/(1+z)$ (1) & 0.299 & 0.0159 \\
\hline
\multicolumn{3}{c} {\em Correlation Tests for Optical Onset Peak Shape (Indices)}  \\
\multicolumn{3}{c} {\em N=20 cases, test not considered accurate for N $<$ 30} \\
 $r$ & -0.603 & 0.0086 \\
 $d$ & -0.173 & 0.4510 \\
 $r/d$ & -0.553 & 0.0159 \\
\hline
\end{tabular}

\medskip
Correlation test results. Variability tested with the peak time includes the limits but excludes the 3 hard-to-classify cases (Figure \ref{f_except}). The peak correlation tests use 73 events and do not support a relationship between the variability and the optical onset time either de-redshifted or as a fraction of the duration $T_{90}$. The 3 hard-to-classify cases were also considered by including their peak time as set by the rise as well as the peak time as set by the decay (see Table \ref{tab_optpkspecial}); neither case changed the results. 
Correlation tests for the peak shape (rising index $r$, decaying index $d$, and their ratio) are included to show that they do not suggest a correlation. Only events with a fitted initial peak where the formal relative uncertainties of indices were $<\, 1/3$ could be considered constrained and their values used. The correlation tests for the indices are not considered accurate due to the small number of available points.
As noted, we considered a background region of 30\% of $T_{90}$ when calculating the variability. We also performed correlation tests with the variabilities calculated using background regions of 10\% and 50\%; this also did not change the results. 
\end{minipage}
\end{table*}

\begin{figure}
\epsscale{1.0}
\plotone{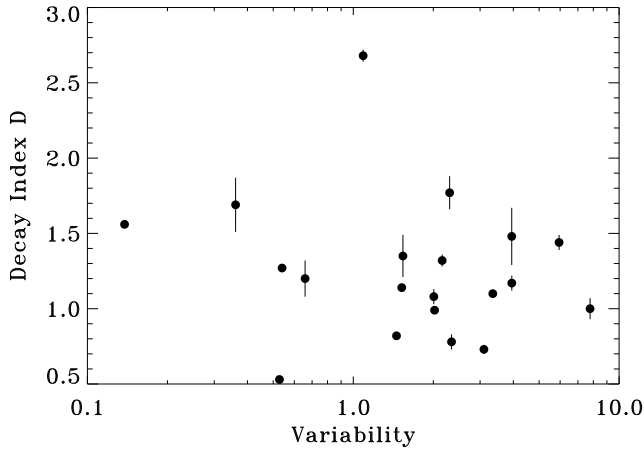}
\caption{Decaying index $d$ vs $\gamma$-ray variability $V$. The indices are from the fit results of Table \ref{tab_optpkfits}, the asymptotic powerlaw of the fitting shape. Only the 20 events where the formal uncertainty is $<\,1/3$ of $d$ are shown. There is no evidence for a connection between the index and the prompt $\gamma$-ray variability. }
\end{figure}

\begin{figure}
\epsscale{1.0}
\plotone{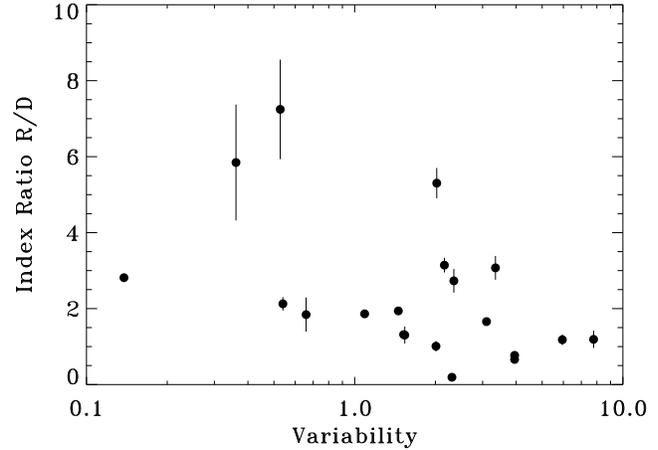}
\caption{Index ratio $r/d$ (rising / decaying) vs $\gamma$-ray variability $V$.  The index ratio is a proxy for the asymmetry in the shape of the optical onset peak. The data suggests a loose relationship,  with the dearth in high-ratio, high-$V$ events. But there is no statistical support for a connection between the properties, with the Spearman test of Table \ref{tab_spear} giving a null probability of 2\% and noting that the 20 events is insufficient to trust the accuracy of the test results. \label{f_ratio}}
\end{figure}

This lack of relationship between the variability $V$ of the prompt GRB and the time it takes to establish the afterglow implies a lack of connection between $V$ and the ``thick'' versus ``thin'' outflow cases of the prompt internal shock model, or the off-axis line of sight or dust destruction timescales in other onset models referred to in the introduction. As noted, the delay is commonly interpreted as indicating the outflow Lorentz factor of a ``thin'' shell. In that case, our result suggests that there would be no strong connection between how variable the GRB outflow is and its overall Lorentz factor.

Other groups have studied the variability of the prompt GRB for any connection to the $\gamma$-ray luminosity, as a luminosity indicator would be an important tool in cosmological studies \citep[e.g.][]{2001ApJ...552...57R, 2007ApJ...660...16S, 2009ApJ...707..387X}. However, \citet{2009ApJ...707..387X} analyzed several prompt GRB properties and noted that $V$ is their ``most noisy'' luminosity relation, compared to properties that included spectral information or energetics; they did not study $V$ as a luminosity indicator in that work. Variability may not be an indicator of the prompt luminosity, yet if it is then our work implies that there is no strong association between the luminosity of the prompt event and the properties of the afterglow onset.

\section{Conclusions}

We compared the optical afterglow onset times (or limits) to the $\gamma$-ray variability $V$ in 76 GRBs with redshifts. In a subset of 25 cases, we fit the shape of the onset ``bump'' as well and compared the rising and decaying indices to $V$.  We did not find any evidence for a pattern and there is no statistical support for any correlations. This indicates a lack of connection between irregularities of the prompt $\gamma$-ray emission and the establishment of the afterglow phase. In the ordinary prompt internal shocks interpretation, this would indicate a lack of relationship between $V$ and the bulk Lorentz factor of the event. 

\section*{Acknowledgements}

We acknowledge the use of public data from the {\em Swift} data archive

We acknowledge the use of data from the ROTSE-III project. The ROTSE project was made possible by grants from NASA, NSF, and the Australian Research Council, and through participation and support from the University of Michigan, Los Alamos National Laboratory, and the University of New South Wales.

STSDAS is a product of the Space Telescope Science Institute, which is operated by AURA for NASA


\bibliographystyle{mn2e}
\bibliography{refs}

%
%
%
%
%
%

\end{document}